\documentclass{aa}
\usepackage{epsfig,slashbox,natbib}
\bibpunct{(}{)}{,}{a}{,}{,}

\begin{document}

\title{EROS~2 proper motion survey: \\
       Constraints on the halo white dwarfs 
\thanks{Based on observations made with the Marly telescope, located 
        at the European Southern Observatory, La Silla, Chile.}}
\author{
B.~Goldman\inst{1,2,3},
C.~Afonso\inst{1,2,3},
Ch.~Alard\inst{4},
J.-N.~Albert\inst{5},
A.Amadon\inst{1},
J.~Andersen\inst{6},
R.~Ansari\inst{5},
{\'E}.~Aubourg\inst{1},
P.~Bareyre\inst{1,2},
F.~Bauer\inst{1},
J.-Ph.~Beaulieu\inst{7},
G.~Blanc\inst{1},
A.~Bouquet\inst{2},
S.~Char\inst{\dag},
X.~Charlot\inst{1},
F.~Couchot\inst{5},
Ch.~Coutures\inst{1},
F.~Derue\inst{1},
R.~Ferlet\inst{7},
P.~Fouqu{\'e}\inst{8,13}, 
J.-F.~Glicenstein\inst{1},
A.~Gould\inst{2,9},
D.~Graff\inst{12},
M.~Gros\inst{1},
J.~Ha\"{\i}ssinski\inst{5},
C.~Hamadache\inst{1},
J.-Ch.~Hamilton\inst{2},
D.~Hardin\inst{10},
J.~de Kat\inst{1},
A.~Kim\inst{1}\thanks{Presently at University of California, Department of Physics, Berkeley, CA 97720, U.S.A.},
Th.~Lasserre\inst{1}, 
L.~Le~Guillou\inst{1}, 
{\'E}.~Lesquoy\inst{1,7},
C.~Loup\inst{7},
Ch.~Magneville \inst{1},
B.~Mansoux\inst{5},
J.-B.~Marquette\inst{7},
{\'E}.~Maurice\inst{11},
A.~Maury\inst{13},
A.~Milsztajn\inst{1}, 
M.~Moniez\inst{5},
N.~Palanque-Delabrouille\inst{1},
O.~Perdereau\inst{5},
L.~Pr{\'e}vot\inst{11},
N.~Regnault\inst{5\star\star},
J.~Rich\inst{1},
M.~Spiro\inst{1},
P.~Tisserand\inst{1},
A.~Vidal-Madjar\inst{7},
L.~Vigroux\inst{1},
S.~Zylberajch\inst{1}
---
The {\sc Eros} collaboration
}
\institute{
CEA, DSM, DAPNIA,
Centre d'{\'E}tudes de Saclay, 91191 Gif-sur-Yvette Cedex, France
\and
Coll{\`e}ge de France, Physique Corpusculaire et Cosmologie, IN2P3 CNRS,
11 pl. M. Berthelot, 75231 Paris Cedex, France
\and
NMSU, Department of Astronomy, Las Cruces, New Mexico 88003, U.S.A.
\and
DASGAL, 77 avenue de l'Observatoire, 75014 Paris, France
\and
Laboratoire de l'Acc{\'e}l{\'e}rateur Lin{\'e}aire,
IN2P3 CNRS, Universit{\'e} Paris-Sud, 91405 Orsay Cedex, France
\and
Astronomical Observatory, Copenhagen University, Juliane Maries Vej 30,
2100 Copenhagen, Denmark
\and
Institut d'Astrophysique de Paris, INSU CNRS,
98~bis Boulevard Arago, 75014 Paris, France
\and
LESIA, Observatoire de Meudon,
92195 Meudon Cedex, France
\and
Ohio State University, Department of Astronomy, 
Columbus, OH 43210, U.S.A.
\and
LPNHE, IN2P3-CNRS-Universit{\'e}s Paris VI et VII, 4 place Jussieu, 
75252 Paris Cedex 05, France
\and
Observatoire de Marseille,
2 pl. Le Verrier, 13248 Marseille Cedex 04, France
\and
University of Michigan, Department of Astronomy, Ann Arbor, MI 48109, U.S.A.
\and 
ESO, Casilla 19001, Santiago 19, Chile
}

\offprints{B.Goldman, bgoldman@nmsu.edu.}

\date{Received; accepted}

\authorrunning{B. Goldman et~al.}
\titlerunning{
  {\sc Eros}~2 proper motion survey:  
  Constraints on the halo white dwarfs
}
\bibliographystyle{aa}

\abstract{
  We are conducting a $377\,^{\circ 2}$ 
  proper motion survey in the $\sim$V and I 
  bands in order to determine the cool white dwarf
  contribution to the Galactic dark matter. Using the 
  $250\,^{\circ 2}$ for which we possess three epochs, and 
  applying selection criteria designed to isolate
  halo-type objects, 
  we find no candidates in a $5500\,{\rm pc}^3$
  effective volume for old, fast $M_V=17$ white dwarfs. 
  We check the detection efficiency
  by cross-matching our catalogue with \citeauthor{Luy79b}'s NLTT catalogue.
  The halo white dwarf contribution cannot exceed 
  5\% (95\% C.L.) for objects with \mbox{$M_V=17$} and \mbox{$1\leq V-I\leq 1.5$}.
  The same conclusion applies to a 14~Gyr halo composed of white dwarfs
  with hydrogen atmosphere, as modeled by \citeauthor{Cha99}.

}
\maketitle

\section{Introduction} \label{intro}

The rotation of the Milky Way is believed to be sustained by a massive dark halo, 
at least beyond one disk scale length. 
Old, cool white dwarfs (WDs) are one of the proposed constituents of 
that dark matter. 
A number of constraints have been placed on the total amount of halo
white dwarfs (HWDs). 
One is based on the light emitted 
by WD~progenitors 
\citep{Gra99}. 
Extragalactic metallicity measurements also place strong constraints
through the metal and helium enrichment due to the HWDs \citep{Fre00}, 
although mechanisms have been proposed to evade those constraints.

Over the past three years a great excitement has come from proper motion 
surveys used to find fast WDs in the Solar neighbourhood.
The surveys conducted by \citet{Luy79} and \citet{Kno99} 
are jointly sensitive to $1500\,{\rm pc}^3$ for $M_V=17.5$ HWDs \citep{Fly01}.
They find no halo WD \citep{Lie88}, 
although this is disputed in the case of \citeauthor{Luy79}'s survey, 
whose efficiency is still a matter of strong debate,
23~years after publication. 
On the other hand, several collaborations 
have reported proper motion detection of fast, cool WDs
some of them being subsequently confirmed spectroscopically
\citep{Iba00,Opp01b}.
\citet{Nel02} also reported the discovery of new HWD candidates in 
the HST Groth--Westphal strip, although using only two epochs.
Finally, \citet{Mon00} discovered new high velocity WDs but the small 
sample of new objects confirms the high efficiency of \citet{Luy79b},
thus decreasing the maximum density of halo WDs.

On the microlensing side, 
\citet{Las00} put a 30\% upper limit 
(95\% C.L.) on the contribution of HWDs of any age
--~or any compact object of similar mass~--
to a standard halo, while \citet{Alc00} 
interpret their 13~microlensing candidates towards the Large
Magellanic Cloud as \mbox{0.15\,--\,$0.9\,{\rm M}_{\mathord\odot}$} 
objects contributing $20^{+11}_{-6}$\% to the standard halo. 

The interesting controversy developed around the \citet{Opp01} results 
addresses the question of the nature of their findings: thick disk or 
halo objects. 
It is in any case agreed that these relatively bright 
$M_{I}\approx 14$, young WDs contribute very little to 
the Galactic dark matter, 
a few percent at most. It has been suggested however that 
these could be the bright tail of a fainter, older population 
\citep{Han01}. 

Our proper motion survey began in 1996 using the EROS 2 cameras. 
It was designed to study the contribution of older,
fainter WDs to the halo, and also addresses the question raised 
by \citet{Han01}. 
Like other proper motion searches, 
ours takes advantage of the large expected velocity dispersion of halo objects,
and it is not intended to address the question of the low velocity, 
or brighter, WDs that most recent surveys probe \citep{Opp01b,Maj02}. 
Here we report our results based on the analysis 
of $250\,^{\circ 2}$ that were observed during at least three epochs.
These yield a sensitivity about three times that of \citet{Luy79b}, 
depending on HWD colours and luminosities.

\section{The EROS~2 proper motion survey} \label{thesurvey}

\subsection{Instrument and observations}  \label{observations}

The {\sc Eros}~2 wide-field imager \citep{Bau98} was 
designed to search for microlensing effects towards the Galactic 
bulge and disk and the Magellanic Clouds.
Its two $1\,^{\circ 2}$ CCD cameras are mounted
at the Cassegrain focus of the 1-m Marly telescope at La Silla
(Chile), with a pixel size of \mbox{$0\arcsec\hskip -5pt .\hskip 2pt 6$}. 
These two $8k\times{}4k$ mosaics are
illuminated through a dichroic beam splitter,
which, together with the CCD efficiency, defines the bandpasses. 
The colour transformations between the {\sc Eros} system 
($\mathcal{V}_{\rm E}$, $\mathcal{I}_{\rm E}$) 
and the standard Johnson-Cousins $(V,I)$ system 
are determined by observing \citet{Lan92} standards
and Ogle \citep{Pac99} secondary standards in the Baade window:
$\mathcal{V}_{\rm E}=V-(0.32\pm 0.03)(V-I)$ and 
$\mathcal{I}_{\rm E}=I+(0.06\pm 0.03)(V-I)$. 
We estimate the precision of the colour transformations to be 
$0.1^{\rm mag}$ for $V-I=\:$0\,--\,2, 
the range of expected colours for HWDs.
Comparison between our $\mathcal{I}_{\rm E}$ photometry 
and I as measured by {\sc Denis} \citep{Fou00}
reveals a small offset, 
\mbox{$I_{\rm Eros}=I_{\rm Denis}-0.06$}, 
while a comparison 
between our photometry and the Photometric GSC \citep{Las88}
reveals no significant offset, with a $0.05^{\rm mag}$ uncertainty. 

Proper motion
observations were performed one to two hours per dark night, 
within 90~minutes of the meridian to minimize atmospheric refraction.
Our limiting magnitudes are about \mbox{$V=21.5$} and \mbox{$I=20.5$}. 
We observed \mbox{$190\,^{\circ 2}$} in the Southern Galactic 
Hemisphere (\mbox{$-79\,^{\circ}<b_{\rm gal}<-48\,^{\circ}$}),
in the following strips along the $\alpha$ coordinate, 
$\Delta \delta = 1.4\,\deg$~wide:
\mbox{22\,h\,16\,min$\,<\alpha<    3\,$h\,44\,min} 
  at \mbox{$\delta=-44\,^{\circ}\,45\,$min},
\mbox{23\,h\,31\,min$\,<\alpha<    1\,$h\,34\,min} 
  at \mbox{$\delta=-40\,^{\circ}\,09\,$min} and
\mbox{22\,h\,24\,min$\,<\alpha\leq 3\,$h\,28\,min} 
  at \mbox{$\delta=-38\,^{\circ}\,45\,$min},
and \mbox{$187\,^{\circ 2}$} in the Northern Hemisphere 
(\mbox{$41\,^{\circ}<b_{\rm gal}<59\,^{\circ}$}) with
\mbox{10\,h\,57\,min$\,<\alpha\leq 13\,$h\,23\,min}
  at \mbox{$\delta=-12\,^{\circ}$} and
\mbox{10\,h\,57\,min$\,<\alpha<    12\,$h\,53\,min}
  at \mbox{$\delta=-4\,^{\circ}\,36\,$min}.

Only those fields ($250\,^{\circ 2}$) with three epochs separated 
by one-year intervals (two-year total baseline),
in one or two bands, are taken into account in this letter.

\subsection{Proper motion catalogue} \label{catalogue}

The reduction software for source detection, classification and
catalogue matching was written in the framework of the {\sc Eros Peida++} 
package \citep{Ans96}. 
Since photon noise dominates the astrometric errors for most of the
search volume, we use a simple two-dimensional Gaussian 
PSF-fitting algorithm 
to determine stellar positions.  A rough star/galaxy classification is
performed to limit galaxy contamination, with cuts chosen so that few stars
are misclassified.  The catalogues of the three epochs are
geometrically aligned, with the deepest one taken as reference,
using a linear transformation fitted to the
40~brightest stars, and matched within a search radius 
corresponding to a $6''{\rm yr}^{-1}$ proper motion. 
The RMS distance between matched stars 
provides an upper limit to the total astrometric error, 
which is 25\,mas (1$\sigma$) for bright objects,
degrading to 150\,mas for $V=21$ or $I=20$. 

In order to minimize contamination by spurious candidates we require 
that all objects be observed at least three times, separated by 
one-year intervals. 
The  proper motions are derived from linear motion fits.
The $\chi^2$ distributions deviate from the expected ones
only for confidence levels lower than 0.5\%, 
indicating the onset of systematic effects.
The 4\% of objects so affected,
mostly poorly measured galaxies, stars close to CCD defects or visual 
binaries,  are removed from the analysis.

\section{Search for halo white dwarfs} \label{search}

\subsection{Selection of halo candidates} \label{selection}

\begin{figure}[bht]
  \begin{center} 
    \epsfig{file=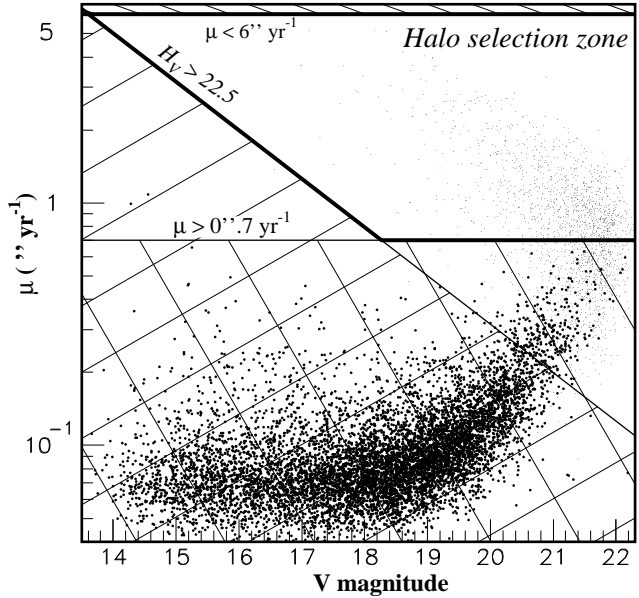} 
    \caption{Proper motion vs. V~magnitude for all objects
      detected in the $\mathcal{V}_{\rm E}$ band,
      whose proper motion significance is higher than 99.7\% C.L.
      (big dots), 
      and for a simulated isotropic standard halo of $M_{V}=17$~WDs
      fully composed of $0.6\,{\rm M}_{\mathord\odot}$~white dwarfs
      (small dots, statistics $\times 10$).
      The selection cuts exclude the hatched area. Here 45\% of the 
      HWDs are lost due to the \mbox{$\mu>0\arcsec\hskip -5pt .\hskip 2pt 7\,{\rm yr}^{-1}$} cut.
    }
    \label{mp}
  \end{center}
\end{figure}

The choice of the criteria used to distinguish halo objects
from the much more numerous disk and thick disk stars, and from 
extragalactic objects, is of prime importance.
We apply two cuts: We first require that the halo candidates 
detected in the $\mathcal{V}_{\rm E}$ band
exhibit a reduced proper motion, 
$H_{V}=V+5\log \frac{\mu}{\rm 1\:''\,yr^{-1}} +5
    =M_{V}+5\log \frac{v_\bot}{\rm 1\:km\,s^{-1}}-3.38>22.5$,
where $\mu$ is the proper motion 
and $v_\bot$ the tangential velocity, 
and that those detected in $\mathcal{I}_{\rm E}$ have $H_{I}>21.5$. 
This cut select candidates that are both 
fast and faint (see Fig.\ref{mp}).

We then require a 
proper motion \mbox{$\mu>0\arcsec\hskip -5pt .\hskip 2pt 7\,{\rm yr}^{-1}$}. 
This cut removes objects 
whose proper motion reality cannot be confirmed by examining our images.
It affects our sensitivity, depending on the mean WD magnitude:
\mbox{$\mu_{\rm min}=0\arcsec\hskip -5pt .\hskip 2pt 7\,{\rm yr}^{-1}$} 
corresponds to a transverse velocity of 
\mbox{$v_\bot=\frac{60\,{\rm pc}}{d}\times 200\,{\rm km\,s}^{-1}$} at distance $d$.
A typical $200\,{\rm km\,s}^{-1}$ transverse velocity is expected 
for halo objects, and $d=60\,$pc is the typical distance of WDs 
in our effective volume for $M_V=17$~stars. 
In fact, a detailed simulation shows that this cut reduces our
sensitivity by 20\% (resp. 45\%)
for $M_V=18$ (resp. 17) HWDs.

The proper motion cut was set 
to \mbox{$\mu>0\arcsec\hskip -5pt .\hskip 2pt 7\,{\rm yr}^{-1}$}
 as a result of the
following optimization: on the one hand this value is low enough
to ensure a high detection efficiency of HWDs and,
 on the other hand, it is sufficiently high to eliminate 
candidates which are likely to be spurious 
because of a poor measurement of their proper motion.
Indeed, when the proper motion cut is lowered 
to \mbox{$0\arcsec\hskip -5pt .\hskip 2pt 6\,{\rm yr}^{-1}$}, 
10 candidates pass the cut, but for 
those whose proper motions were measured in both bands (6 among the
10 candidates) we find that for all of them the measured values 
of their visible and red proper motion are incompatible.

No candidates survive the two cuts: $H_V>22.5$ or \mbox{$H_I>21.5$},
and \mbox{$\mu>0\arcsec\hskip -5pt .\hskip 2pt 7\,{\rm yr}^{-1}$}. 
Had we required \mbox{$\mu>0\arcsec\hskip -5pt .\hskip 2pt 8\,{\rm yr}^{-1}$}, 
our upper limits on the HWD contribution to the halo given below 
would move up by less than 1\% of the standard dark halo.

\subsection{Comparison with \citeauthor{Luy79b}'s catalogue} \label{efficiency}

We cross match our proper motion catalogue 
with the NLTT catalogue \citep{Luy79b},
by searching for bright stars in our catalogue 
with proper motion higher 
than \mbox{$0\arcsec\hskip -5pt .\hskip 2pt 1\,{\rm yr}^{-1}$}, 
within 1' of Luyten positions. 
It turns out that 85\% of  \citeauthor{Luy79b}'s stars are recovered with compatible proper motion 
and proper motion angle, 
which demonstrates our high detection 
efficiency for stars similar to \citeauthor{Luy79b}'s stars.
The 15\% loss is partly due to the expected loss due to defects in the
CCDs (4\%) and to pointing dispersion between epochs (4\%).
Additionally, our proper motion measurements for \citeauthor{Luy79b}'s stars are 
significant at the $3\sigma$ level in each band, and are compatible 
at the $2\sigma$ level. This gives us confidence in our proper motion 
accuracy, as most Luyten stars have proper motion much lower than our 
\mbox{$0\arcsec\hskip -5pt .\hskip 2pt 7\,{\rm yr}^{-1}$} halo selection cut.
However, this only applies to bright stars, as there are no Luyten 
stars close to our photometric detection limit.
Unfortunately we can draw no conclusion regarding  \citeauthor{Luy79b}'s completeness 
based on this study, as we lack a reliable estimate of the contamination 
of our disk-like proper motion catalogue.
This study will be reported in more detail in a forthcoming article.

\section{Discussion} \label{discussion}

\subsection{Halo model predictions} \label{predictions}

In order to check our sensitivity to HWDs, we perform a full
simulation of our observations. We measure the detection efficiency
of a star on each frame by adding simulated stars to real images. 
Using the Besan{\c c}on model of the Galaxy \citep{Rob00},
to which we add one HWD per field,
we then construct a simulate catalogue for each of our fields and
each of our images, and process these catalogues 
in the same way as our observations.
The Besan{\c c}on model is a consistent hydrodynamical description 
of Galactic star counts and proper motion,
based on initial mass function and stellar evolution.

We use various halo models with different kinematics.
We use isotropic velocity dispersions 
  of $\sigma_{\rm 1D}=100$, 130 or $156\,{\rm km\,s}^{-1}$,
or anisotropic, spheroid-like 
dispersions, with $\sigma_{\rm 1D}=85$, 
105 or $125\,{\rm km\,s}^{-1}$ 
along the $V$ (Galactic rotation) direction.
We suppose halo rotations of $-50$, 0 or $50\,{\rm km\,s}^{-1}$. 
The expected number of detections depend on these parameters 
at the 5 to 10\% level. 
For a given kinematics, flattened halos correspond to higher local 
densities, and thus to proportionately more detections.

We consider HWDs of \mbox{$16.5\leq M_V\leq 18$} 
and \mbox{$-0.5\leq V-I\leq 1.5$}, in steps of 0.5\,mag.
We also generate HWDs with hydrogen atmosphere 
according to the \citet{Cha99} luminosity
functions for halo ages of 14 and 15~Gyr. 
The predicted number of detections, in either the $\mathcal{V}_{\rm E}$ 
or the $\mathcal{I}_{\rm E}$ band, 
are reported in Table~\ref{expect}, for the standard isotropic,
isothermal halo, which has a local density of 
$8.10^{-3}\,{\rm M}_{\mathord\odot}\,{\rm pc}^{-3}$.
Statistical errors are 5\% 
while systematic errors are estimated to be less than 25\%, 
the quadrature sum of errors in 
calibration (15\%), efficiencies (10\%)
and parameters of the halo kinematics (15\%).

\begin{table}[hbt]
  \caption{
    Number of detected WDs, 
    in either $\mathcal{V}_{\rm E}$ or $\mathcal{I}_{\rm E}$ band, 
    predicted for a standard, isotropic halo with a local HWD mass density of 
    $8.10^{-3}\,{\rm M}_{\mathord\odot}\,{\rm pc}^{-3}$,
    and for $0.6\,{\rm M}_{\mathord\odot}$~HWDs 
    of various colours and magnitudes, over $250\,^{\circ 2}$.
    The effective volume is given between parenthesis for $M_V=17$ 
    in units of $10^3\,{\rm pc}^3$.
    \label{expect}}
  \begin{center}
    \begin{tabular}{|c|r|rr|r|r|}
      \hline
      \backslashbox{$V-I$}{$M_V$} & 16.5 
       & \multicolumn{2}{|c|}{17} & \multicolumn{1}{|c|}{17.5}
       & \multicolumn{1}{|c|}{18} \\
      \hline                   
      $-0.5$ &  48.2 &  35.9 & (2.69) & 16.7 & 9.8  \\ 
      0.     &  69.2 &  38.2 & (2.87) & 24.5 & 14.7  \\ 
      +0.5   &  83.9 &  46.5 & (3.49) & 30.9 & 17.3 \\
      1.     & 106.5 &  74.1 & (5.56) & 40.0 & 21.4 \\
      1.5    & 140.6 & 111.3 & (8.35) & 58.7 & 29.2 \\
      \hline
    \end{tabular}
  \end{center}
\end{table}

We have not taken binarity effects into account.
Double stars of similar magnitude would appear brighter than
individual WDs, so that objects at the brighter, slower end of the
distribution could  be lost by the reduced proper motion cut.
Furthermore, the number of systems to which we are sensitive 
would be smaller than the actual number of stars; 
our constraints would have to be relaxed accordingly.

\subsection{Constraints on the halo 
  WD fraction and age} 
           \label{constraints}

\begin{figure}[th]
  \begin{center} 
    \epsfig{file=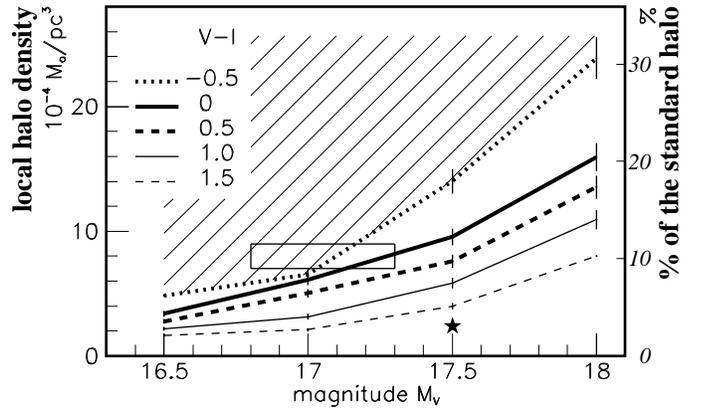} 
    \caption{Constraints on the WD contribution to the standard halo,
      for various WD colours, at the 95\%~C.L. 
      The error bars correspond to the statistical error in the 
      simulations. We exclude the hatched area.
      The empty box 
      corresponds to the two WDs detected by 
      \citet{Iba00}, the star the 
      result by \citet{Fly01} (no error bar given).
    }
    \label{const}
  \end{center}
\end{figure}

The constraints on the contribution of WD to the halo are obtained from 
the above predictions and the lack of halo candidates, 
and shown in Fig.~\ref{const}.
We find that a 14~Gyr~halo cannot be made of more 
than 5\% 0.6 
or $1.0\,{\rm M}_{\mathord\odot}$ HWDs, at the 95\% C.L.,  
while a 15~Gyr~halo cannot be made of more 
than 15\% of $0.6\,{\rm M}_{\mathord\odot}$ HWDs, 
  or 45\% of $1.0\,{\rm M}_{\mathord\odot}$ HWDs. 

\subsection{Comparison with other surveys} \label{comparison}

Our results, based on a larger volume than the study by \citet{Fly01}, 
confirm their conclusion.
They are marginally compatible with 
results of \citet{Iba00}, depending
on the exact WD colours and absolute magnitudes.

Regarding \citet{Opp01b}, 
we do not exclude a 1~to 2\% contribution by bright $M_I$=13\,--\,15 HWDs.
We stress that our selection cut, 
\mbox{$\mu>0\arcsec\hskip -5pt .\hskip 2pt 7\,{\rm yr}^{-1}$}, 
is more severe than theirs, while preserving most of the halo sensitivity
{\it for the faint magnitudes} of interest here.
\citet{Rei01}, \citet{Rey01} and \citet{Koo02}
also discuss this survey.
We do constrain the proposition of \citep{Han01} 
that the WDs detected by \citet{Opp01b} be the 
bright tail of a fainter population. 
If this population has a halo-like kinematics, 
then the older part of the luminosity function, 
older by several Gyrs, cannot make up more than $\approx 7\%$ of the halo.

Finally, our constraints are fully compatible with the {\sc Eros}
microlensing results. 
The constraints from the present study are even stronger than those 
from microlensing, for HWDs of absolute 
magnitude $M_V<18$ and also if the halo age is smaller than 15~Gyr 
(using the \citet{Cha99} luminosity function for DAs). 

\section{Conclusion}

In this letter we have presented the analysis of $250\,^{\circ 2}$ of the
{\sc Eros} proper motion survey, dedicated to the search of old, 
faint white dwarfs. We found no halo white dwarf candidates. 
We exclude a white dwarf contribution to the halo above 
5\% for $M_V=17$, $1\leq V-I\leq 1.5$, at the 95\% C.L.
Assuming a \citet{Cha99} cooling model, a 14~Gyr halo cannot be made of 
more than 5\% of WDs with hydrogen atmosphere.
Our limits degrade to 20\% for a 15~Gyr halo 
of $1\,{\rm M}_{\mathord\odot}$ white dwarfs 
or for $M_V=18$, $0\leq V-I\leq 1.5$ white dwarfs.

\begin{acknowledgements}
We thank Annie Robin and her collaborators 
for allowing us to use the latest version of the Besan{\c c}on model
prior to publication. 
We thank also Nicolas Epchtein and the {\sc Denis} collaboration for 
providing infrared unpublished photometry of our sources in order to 
check our $\mathcal{I}_{\rm E}$ calibration, and Gilles Chabrier for 
his comments.
This research has made use of the {\sc Simbad} database, operated at 
CDS, Strasbourg, France. 
We are grateful to Daniel Lacroix and the technical staff at the
Observatoire de Haute Provence and to Andr{\'e} Baranne for their help in
refurbishing the MARLY telescope and remounting it in La Silla. We
also thank the technical staff of ESO La Silla for their support of
{\sc Eros}. We thank Jean-Fran{\c c}ois Lecointe and Adelino Gomes
for assistance with the online computing. 
Work by A.G. was supported by NSF grant AST~97-27520 and by
a grant from le Centre fran{\c c}ais pour l'accueil et les {\'e}changes
internationaux.
\end{acknowledgements}

\end{document}